\documentclass[aps,pra,english,twocolumn,letterpaper,notitlepage]{revtex4-2}

\usepackage[T1]{fontenc}
\usepackage{graphicx}
\usepackage{epstopdf}
\usepackage{amssymb}
\usepackage{amsmath}
\usepackage[urlcolor=blue,hyperindex,colorlinks,bookmarks=true,linkcolor=red,citecolor=black]{hyperref}
\usepackage[normalem]{ulem}
\usepackage{color}
\usepackage{rotating}
\usepackage{physics}
\usepackage{bm}
\usepackage{soul}
\usepackage{enumerate}

\usepackage[english]{babel}

\def\iden{\hat{\mathbb{I}}}

\newcommand{\superket}[1]{\left|\left.#1\right>\hspace{-2pt}\right>}
\newcommand{\superbra}[1]{\left<\hspace{-2pt}\left<#1\right.\right|}

\begin{document}

\title{Nonlinear input transformations are ubiquitous in quantum reservoir computing}

\author{L.~C.~G.~Govia}
\email{lcggovia@gmail.com}
\author{G.~J.~Ribeill}
\author{G.~E.~Rowlands}
\author{T.~A.~Ohki}
\affiliation{Quantum Engineering and Computing, Raytheon BBN Technologies, 10 Moulton St., Cambridge, MA 02138, USA}

\begin{abstract}
 The nascent computational paradigm of quantum reservoir computing presents an attractive use of near-term, noisy-intermediate-scale quantum processors. To understand the potential power and use cases of quantum reservoir computing, it is necessary to define a conceptual framework to separate its constituent components and determine their impacts on performance. In this manuscript, we utilize such a framework to isolate the input encoding component of contemporary quantum reservoir computing schemes. We find that across the majority of schemes the input encoding implements a nonlinear transformation on the input data. As nonlinearity is known to be a key computational resource in reservoir computing, this calls into question the necessity and function of further, post-input, processing. Our findings will impact the design of future quantum reservoirs, as well as the interpretation of results and fair comparison between proposed designs.
\end{abstract}

\maketitle

\section{Introduction}

Quantum reservoir computing (QRC) is an emerging area in the field of quantum neuromorphic computing \cite{Markovic:2020aa,Mujal21} that promises the potential application of near-term quantum technology to real-world problems. It generalizes the classical computing paradigm of reservoir computing \cite{Maass:2002aa,Jaeger:2004aa,Verstraeten:2007aa} to consider quantum systems as the computational resource, or reservoir. Unlike in traditional machine learning approaches, such as neural networks, the internal dynamics of the reservoir are untrained in this paradigm. Instead, to achieve task performance the reservoir response to a time-dependent signal is recorded, and post-processed using a linear weight-matrix. This weight matrix is the only trainable element of the setup.

Considerable effort has been devoted to both understanding what makes a high performing reservoir, and using this understanding to design physical reservoirs that implement such dynamics naturally \cite{Tanaka:2019aa}. To that end, quantum systems appear attractive due to i) their naturally large state-space available for computation, ii) potential for stronger-than-classical correlations, and, as we shall see, iii) ubiquitous nonlinearity. It is an active area of research to understand the conditions necessary to access the full benefit of quantum reservoirs.

In designing computational reservoirs, it is useful to elucidate which parts of the reservoir dynamics are necessary for performance, both to reduce useless complexity, and to understand limiting factors. In light of this, it is important to understand the nature of the dynamical map of the input data into reservoir state variables, which we refer to as the input encoding. It has been shown for classical reservoirs that a nonlinear input encoding may be sufficient for task performance, even if the rest of the dynamics are fully linear \cite{Abreu-Araujo:2020wy}.

In this manuscript, we explore the nature of input encodings in QRC for explicitly time-dependent tasks, and find that by virtue of quantum mechanics nonlinear input encodings are ubiquitous. We consider both discrete encodings, where time-series data is discretized and feed-in sequentially, and continuous encodings, where the input is simultaneous with internal reservoir dynamics and measurement. In both cases, we derive a set of conditions under which the input encoding is linear. For the purposes of this discussion, we will consider the computational nodes of a quantum reservoir to consist of the expectation values of a complete basis for the operator space acting on the reservoir's Hilbert space. We will often, though not exclusively, consider bases that consist of Hermitian matrices, i.e.~observables. 

This manuscript is organized as follows. In section \ref{sec:encode} we formalize our definition of a linear input encoding, and derive the conditions required for a linear encoding in QRC. In sections \ref{sec:finite} and \ref{sec:infinite} we outline examples of input encodings for finite and infinite dimensional quantum reservoirs found in the literature, and discuss their (non)linearity. Finally, in section \ref{sec:conc} we make our concluding remarks.

\section{Input encodings}
\label{sec:encode}

For an intrinsically time-dependent, $N$-dimensional input signal vector $\mathbf{u}(t)$, we separate QRC on time-dependent tasks into two classes:
\begin{enumerate}
    \item \textbf{Discrete input encoding}: $\mathbf{u}(t)$ is discretized into a finite series of vectors $\mathbf{u}_j$, where $j$ indexes the time-step, which are sequentially input into the reservoir. The approaches within this class typically have distinct input, reservoir internal evolution, and measurement steps, described by separate quantum channels.
    
    \item \textbf{Continuous input encoding}: the time dependent signal $\mathbf{u}(t)$ is feed directly into the system in a continuous fashion, concurrent with internal system evolution and measurement.
\end{enumerate}

\begin{figure*}
    \centering
    \includegraphics[width=1.9\columnwidth]{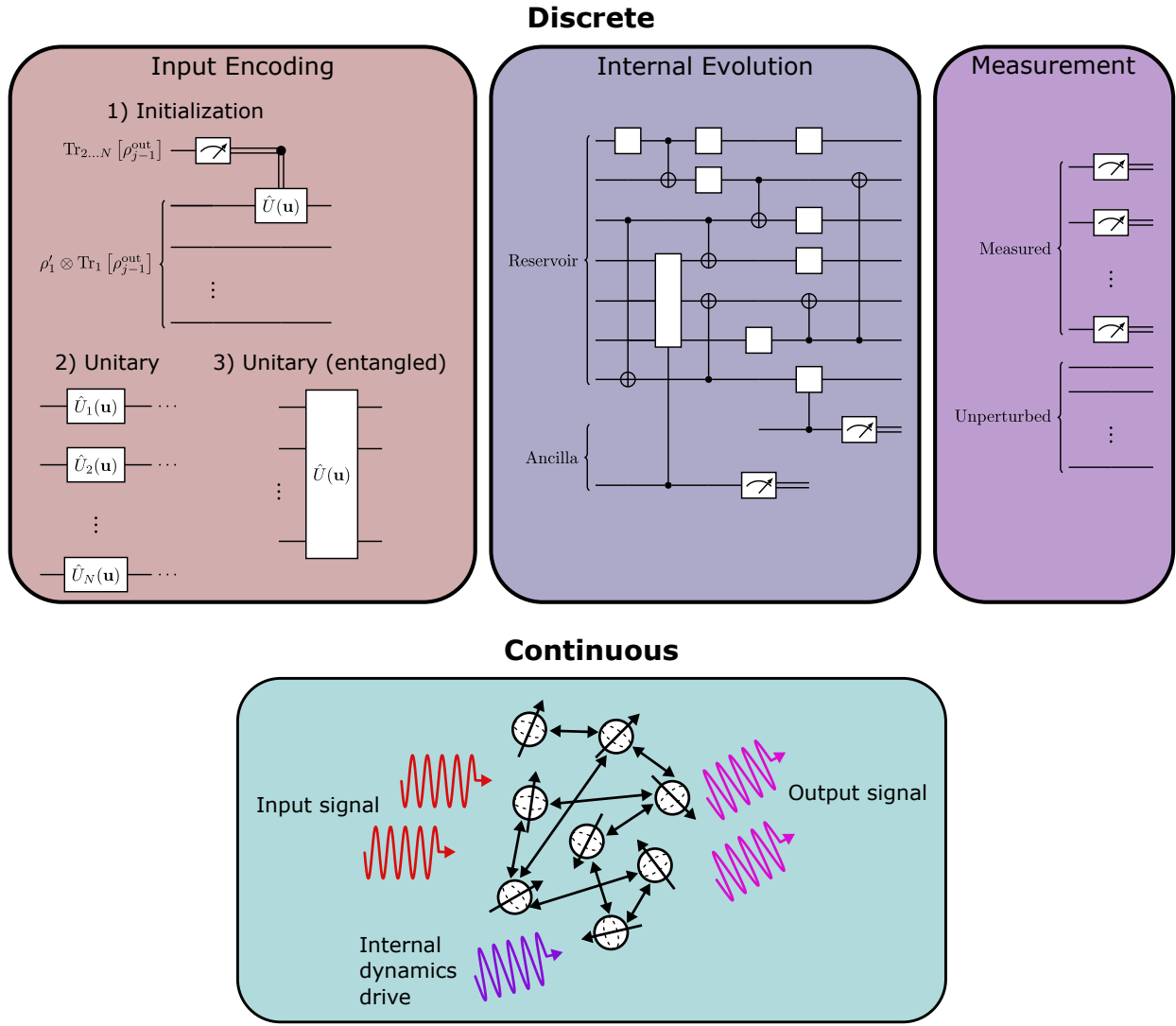}
    \caption{Quantum reservoir computing with a discrete (upper panel) or continuous (lower panel) input encoding. ({\bf Upper panel}) Example quantum circuits for the input encoding, including state (re)initialization and unitary encodings; the internal reservoir dynamics, which may make use of ancillary systems to implement incoherent dynamics; and the measurement step. ({\bf Lower panel}) Continuous-time quantum reservoirs have simultaneous input, internal dynamics, and measurement, represented by the control fields incedent on the reservoir (e.g.~a network of spins as shown). }
    \label{fig:schemes}
\end{figure*}

We consider an input encoding linear if for any input time ($t_j$ for the discrete case and $t$ for continuous) the expectation value of any reservoir system operator (i.e.~the state of the computational node) is a linear function of the input signal at that time. Explicitly, for discrete and continuous input encodings we require
\begin{align}
    &\text{Discrete:}~\left<\hat{B}_k(t_j)\right> \propto f_k^{1}(\mathbf{u}_j) + ... \\
    &\text{Continuous:}~\left<\hat{B}_k(t)\right> \propto f_k^{1}(\mathbf{u}(t)) + ... \label{eqn:linearenc}
\end{align}
for all elements $\hat{B}_k$ of an operator basis for the reservoir system, indexed by $k$, where $f^{1}: \mathbb{R}^{N} \rightarrow \mathbb{C}$ is a linear function of the elements of $\mathbf{u}$ that returns a single complex number.

By definition, a linear encoding in one basis implies a linear encoding in all other bases, since they are connected by linear transformations. Note that the other terms of Eq.~\eqref{eqn:linearenc}, encapsulated in ``$+...$'', can include nonlinear functions of the previous state of the reservoir nodes, and therefore of the input history, i.e.~$\mathbf{u}_l$ for $l < j$ or $\mathbf{u}(s)$ for $s < t$. While likely necessary for reservoir computational performance, this does not impact the discussion of the linearity of the input.

An important point that we emphasize in this discussion is that we require that we achieve input-relations of the form of Eq.~\eqref{eqn:linearenc} without nonlinear pre-processing of $\mathbf{u}(t)$. The reason for this is twofold. Firstly, with arbitrary nonlinear pre-processing it is always possible \emph{in principle} to achieve a linear input encoding. This renders our entire discussion pointless, since we are interested in the conditions on QRC that allow for a linear input encoding. Secondly, nonlinear pre-processing is a computational resource in itself, and may alone be sufficient as the resource in a reservoir computing framework. Thus, we argue that allowing for nonlinear pre-processing is in opposition to the goal of understanding the computational power of QRC.

\subsection{Discrete input encodings}

The total evolution of the state during one time-step of a discrete input encoding QRC protocol is given by
\begin{align}
    \rho_j = \mathcal{C}_{\rm out}\left[\mathcal{C}_{\rm res}\left[\mathcal{C}_{\rm in}(\mathbf{u}_j)\left[\rho_{j-1}\right]\right]\right],
\end{align}
where $C_{\mu}$ for $\mu\in\{{\rm in},~{\rm res},~{\rm out}\}$ are the quantum channels describing the input, evolution, and measurement steps. QRC operation of this form is similar to the implementation of classical reservoir computing in software, such as echo state networks \cite{Jaeger:2004aa}. We note that for the purposes of this manuscript the internal dynamics and measurement can happen simultaneously. Only the input step must be separated for the discussion that follows.

The benefits of this approach are that it aligns more closely to the discrete circuit-model picture of quantum computation, and it can be used to artificially slow down the data input-rate. The main drawbacks are that it does not take full advantage of the time-dependent nature of a reservoir built from a physical dynamical system in hardware, nor does it allow for processing at the (potentially fast) natural timescale of the physical reservoir.

As discussed in Ref.~\cite{Chen:2019aa}, all discrete input encodings can be described by a quantum channel that is parameterized by the elements of $\mathbf{u}_j$, which we write as 
\begin{align}
    \rho^{\rm in}_j = \mathcal{C}_{\rm in}(\mathbf{u}_j)\left[\rho_{j-1}\right].
\end{align}
The linear encoding condition for discrete input can then be expressed as
\begin{align}
    {\rm Tr}\left[\hat{B}_k \rho^{\rm in}_j\right] = f_{k,j}^{1}(\mathbf{u}_j), \label{eqn:linearencdiscrete}
\end{align}
for a basis of operator space $\{\hat{B}_k\}$.

From the above expression, it is clear that the requirements of a linear encoding depend entirely on the form of the state $\rho^{\rm in}_j$ after the input step of each time-step, and can simply be expressed by the requirement that in any basis decomposition of $\rho^{\rm in}_j$ the coefficients are linear functions of the elements of $\mathbf{u}_j$. That is to say we require that
\begin{align}
    \rho^{\rm in}_j = \sum_m f_{m,j}^{1}(\mathbf{u}_j) \tau_m, \label{eqn:fintiediscrete}
\end{align}
for $\{\tau_m\}$ a basis of operator space. Taking $\{\tau_m\}$ to be an orthogonal basis, it is clear that the linear encoding condition of Eq.~\eqref{eqn:linearencdiscrete} is valid if and only if Eq.~\eqref{eqn:fintiediscrete} is valid. As a reminder, if Eq.~\eqref{eqn:fintiediscrete} is true for one basis it is true for all by linearity of basis transformations.

While seemingly benign, the linear encoding condition is actually a strong constraint on the nature of the input channel, since Eq.~\eqref{eqn:fintiediscrete} must be valid for \emph{any} state $\rho_{j-1}$ of the previous step. Later in this manuscript we will explore examples of discrete encodings present in the literature, some of which will be linear, and some not. We leave a detailed understanding of the form of channels that satisfy linear encoding for future work.

\subsection{Continuous input encodings}

We define a continuous-mode quantum reservoir to be a dynamical quantum system that experiences a continuous force or ``drive'' that is a function of the input signal $\mathbf{u}(t)$. In this operation mode, there are no longer distinct input, reservoir evolution, and measurement steps, but rather all three happen simultaneously. The total evolution of the reservoir can therefore be described by the parameterized quantum channel
\begin{align}
    \rho(t) = \mathcal{C}(\mathbf{u}(t),\boldsymbol{\alpha})[\rho(t)],
\end{align}
where $\boldsymbol{\alpha}$ are the hyper-parameters of the reservoir dynamics and measurement that are independent of the input. 

We will focus on quantum channels that can be described in differential form by a Lindbladian generator
\begin{align}
    \partial_t\rho(t) = \mathcal{L}(\mathbf{u}(t),\boldsymbol{\alpha})[\rho(t)],
\end{align}
as this allow us to write the equivalent adjoint master equation for the evolution of operators
\begin{align}
    \partial_t\hat{B}(t) = \mathcal{L}^\dagger(\mathbf{u}(t),\boldsymbol{\alpha})[\hat{B}(t)], \label{eqn:diffmapfinite}
\end{align}
where $\mathcal{L}^\dagger$ is the Hermitian adjoint operator of $\mathcal{L}$. Furthermore, by focusing on the differential map, we align ourselves more closely with the mindset of the operation of physical devices, where one often thinks of the evolution in terms of controllable Hamiltonian (or even dissipative) dynamics, as opposed to the integrated channel.

However, we must be careful when describing the reservoir dynamics in differential form, as it may be tempting to assume that the input encoding is linear if $\mathcal{L}^\dagger$ depends linearly on the elements of $\mathbf{u}(t)$. This is not the case, as from Eq.~\eqref{eqn:linearenc} we see that linearity of the encoding will depend on the integration of Eq.~\eqref{eqn:diffmapfinite}. 

We can always write Eq.~\eqref{eqn:diffmapfinite} (restricted to the input-dependent parts) in an operator basis as
\begin{align}
    \partial_t\hat{B}(t) = \sum_{k} f_k(\mathbf{u}(t)) \hat{B}_{k}
\end{align}
where for reasons that will soon be apparent, we fix $\hat{B}_1 = \iden$ the identity operator. Even if all $f_k$ are linear functions of $\mathbf{u}(t)$, if any terms of the right-hand side of the above equation are nonzero other than that for the identity matrix (i.e.~$k=1$), the solution for $\hat{B}(t)$ will be defined by an exponential map. To see this, we set $\hat{B} = \hat{B}_n$ an element of the basis, and then for the vector of expectation values of the basis elements
\begin{align}
    \mathbf{B} = \left(\left<\hat{B}_1\right>,\left<\hat{B}_2\right>,...\right)^\intercal, \label{eqn:expvec}
\end{align}
we can write the differential equation
\begin{align}
   \partial_t \mathbf{B}(t) = \mathbf{M}(\mathbf{u}(t))\mathbf{B}(t), \label{eqn:matvecdif}
\end{align}
where $\mathbf{M}(\mathbf{u}(t))$ is a matrix that expresses the differential equations 
\begin{align}
    \partial_t\left<\hat{B}_n(t)\right> = \sum_{k} f_{k,n}(\mathbf{u}(t)) \left<\hat{B}_{k}\right>,
\end{align}
for all $n$. 

The solution to Eq.~\eqref{eqn:matvecdif} can be expressed as $\mathbf{B}(t) = \exp[\mathbf{P}(\mathbf{u}(t))]\mathbf{B}(0)$, where $\mathbf{P}(\mathbf{u}(t))$ will often have no closed form solution. In the most general case where $\mathbf{M}(\mathbf{u}(t))$ does not commute with itself at different times, a series expansion for $\mathbf{P}(\mathbf{u}(t))$ known as the Magnus expansion can be constructed \cite{Magnus1954}, and the lowest order term in this series is intuitively given by
\begin{align}
    \mathbf{P}^{(1)}(\mathbf{u}(t)) = \int_0^t \mathbf{M}(\mathbf{u}(s)) ds.
\end{align}
Regardless of the form of $\mathbf{P}(\mathbf{u}(t))$, it is clear that an exponential map has been applied to the matrix $\mathbf{M}(\mathbf{u}(t))$. Thus, the input encoding of $\mathbf{u}(t)$ cannot be linear unless $f_{k,n}(\mathbf{u}(t))$ are chosen precisely to invert the matrix exponential. This is generally impossible, since $f_{k,n}(\mathbf{u}(t))$ act element-wise in $\mathbf{M}(\mathbf{u}(t))$ and the matrix exponential does not, but in any case would require a nonlinear pre-processing step applied to $\mathbf{u}(t)$.

The only way to avoid this is to require that the input part of the differential equation governing the reservoir dynamics is of the form
\begin{align}
    \partial_t\hat{B}_n(t) = f_{1,n}(\mathbf{u}(t))\iden,
\end{align}
for all elements $n$ of an operator basis where the first element is the identity. It is straightforward to integrate this to obtain the solution for the expectation value
\begin{align}
    \left<\hat{B}_n(t)\right> = \int_{t_0}^t ds f_{1,n}(\mathbf{u}(s)), \label{eqn:defint}
\end{align}
which is \emph{still} not guaranteed to be a linear encoding, since the definite integral will depend on the functional form of $f_{1,n}(\mathbf{u}(s))$. 

If $f_{1,n}$ is an arbitrarily weighted summation of the elements of the derivative of $\mathbf{u}(t)$ with respect to time, then a linear input encoding is obtained. This can be approximated to a high degree of accuracy by a linear function that has access to $\mathbf{u}(t)$ at times in the past and future (using numerical differentiation). While such pre-processing is strictly speaking linear, it is also something we consider outside of the scope of acceptable pre-processing because it requires $\mathbf{u}(s)$ for $s > t$ to calculate. 


To recover a linear input encoding without complex pre-processing, we remind ourselves that to ensure the fading memory property any functioning reservoir will have finite lifetimes $1/\gamma_n$ for all basis elements. Adding this to the differential equation we have
\begin{align}
    \partial_t\left<\hat{B}_n(t)\right> = f_{1,n}(\mathbf{u}(t)) - \gamma_j \left<\hat{B}_n(t)\right>,
\end{align}
which has the general solution
\begin{align}
   \nonumber \gamma_n\left<\hat{B}_n(t)\right> &= f_{1,n}(\mathbf{u}(t)) - e^{-\gamma_n(t-t_0)}f_{1,n}(\mathbf{u}(t_0)) \\ &+ \int_{t_0}^t ds~e^{-\gamma_n(t-s)} C_n(\mathbf{u}(s), s), \label{eqn:intexplicit}
\end{align}
where $C_n(\mathbf{u}(t), t) = \frac{d}{dt}f_{1,n}(\mathbf{u}(t))$ is the full derivative of $f_{1,n}(\mathbf{u}(t))$ with respect to time. The first line of the above equation has the form of a linear input encoding, provided $f_{1,n}$ is a linear function, with an exponentially decaying contribution form the initial input. 

The second line of Eq.~\eqref{eqn:intexplicit} complicates matters. Expanding the integration kernel $C_n(\mathbf{u}(t), t)$ under the assumption that $f_{1,n}(\mathbf{u}(t))$ is a linear function of $\mathbf{u}(t)$ we have
\begin{align}
   \nonumber C_n(\mathbf{u}(t), t) &= \partial_t f_{1,n}(\mathbf{u}(t)) + \nabla_{\mathbf{u}}f_{1,n}(\mathbf{u}(t)).\partial_t \mathbf{u}(t) \\
   &= \sum_l u_l(t) \partial_t c_{n,l}(t) + \sum_l c_{n,l}(t) \partial_t u_l(t),
\end{align}
where $f_{1,n}(\mathbf{u}(t)) = \sum_l c_{n,l}(t) u_l(t) $ is the general form for a linear function of the elements $u_l(t) = [\mathbf{u}(t)]_l$. Assuming $c_{n,l}(t) = c_{n,l}$, which in reservoir computing language implies that the input weights or mask are time-independent, we see that the input encoding of Eq.~\eqref{eqn:intexplicit} is linear except for the nonlinear contribution
\begin{align}
    \left<\hat{B}_n(t)\right>_{\rm NL} = \frac{1}{\gamma_n}\sum_l c_{n,l}\int_{t_0}^t ds~e^{-\gamma_n(t-s)}\partial_s u_l(s).
\end{align}
Thus, the input encoding is linear if the derivative of the input signal vanishes, as is the case for constant input. Otherwise, there will be a nonlinear contribution to the input encoding.

To summarize, we have arrived at two conditions to obtain a linear continuous input encoding:
\begin{enumerate}[i)]
    \item The input acts as a linearly encoded inhomogeneous forcing term: $\partial_t\hat{B}_n(t) = f^1_{1,n}(\mathbf{u}(t))\iden + ...$
    
    \item The input signal $\mathbf{u}(t)$ is constant or piece-wise constant.
\end{enumerate}
In regards to criteria ii), if one ignores the non-zero derivative at the edge of piece-wise constant sections, then the encoding is linear. This can be justified by the fact that any physical implementation of continuous input will have finite resolution, such that one can treat piece-wise constant input as a series of constant input evolutions with updating initial conditions. This brings continuous input closer to discrete, but the internal dynamics and measurement still happen simultaneously.

As we will show with examples in the rest of this manuscript, only a highly select set of QRC models will satisfy both of these conditions. Note that in Appendix \ref{app:gencon} we explain why even when we consider in full generality the internal dynamics of the reservoir that occur simultaneously with the input, the above conditions are unmodified.

\section{Finite dimensional systems}
\label{sec:finite}

We first consider quantum reservoirs with a finite dimensional Hilbert space, for example, a single qudit in a $d$-dimensional Hilbert space, or an array of qubits. We separate the following examples into discrete and continuous encodings.

\subsection{Discrete input encodings}

\subsubsection{State re-initialization}

The first example of a discrete input encoding is a re-initialization of the state of one (or more) sub-components of the full system, such that the state after input at step $j$ is given by
\begin{align}
    \rho^{\rm in}_j = \bigotimes_{n=1}^{P_{\rm in}}\sigma_n(\mathbf{u}_j) \otimes {\rm Tr}_{1...P_{\rm in}}\left[\rho_{j-1}\right],
\end{align}
where $P_{\rm in}$ is the number of sub-components used for input. This is the input encoding employed in a majority of QRC studies to date \cite{Fujii:2017aa,Nakajima:2019aa,Negoro:2018aa,Kutvonen:2020aa,Martinez-Pena:2020ts,Fujii2020,Pena2021,Xia2021,Tran2021}. The linearity of this input encoding depends on the linearity (with respect to the input parameters) of the states $\sigma_n$ themselves. A linear encoding will satisfy
\begin{align}
    {\rm Tr}\left[\hat{B}_k\sigma_n\right] = f_{k,n,j}^1(\mathbf{u}_j),
\end{align}
for all elements of the basis and all $\sigma_n$, where the specific linear function may depend on the basis element $k$, the state $n$, and the time-step $j$. 

As an example, consider the scheme first introduced in Ref.~\cite{Fujii:2017aa}, which re-initializes a single qubit at each time-step via
\begin{align}
    \ket{\sigma_j} = \sqrt{1-u_j}\ket{0} + \sqrt{u_j}\ket{1}.
\end{align}
It is easy to show that while the encoding is linear in the Pauli-Z operator, as $\left<\hat{Z}\right> = 1 - 2u_j$, it is nonlinear in the Pauli-X operator: $\left<\hat{X}\right> = \sqrt{u_j(1-u_j)}$. A minimal modification that would create a fully linear encoding would be an input scheme with no coherence, such as 
\begin{align}
    \sigma_j = (1-u_j)\ketbra{0} + u_j\ketbra{1},
\end{align}
which was used in Refs.~\cite{Chen:2019aa,Tran2020}.

More generally, coherent superpositions where the coefficients depend on the elements of $\mathbf{u}_j$ as
\begin{align}
    \ket{\sigma_n} = \sum_m \sqrt{f_{m,n,j}(\mathbf{u}_j)} \ket{\psi_m}
\end{align}
result in a square-root nonlinearity in the expectation values of the operators $\ketbra{\psi_m}{\psi_l}$ for $m\neq l$. Re-initialization input encodings are linear if they can be written as convex sums of the form
\begin{align}
    \bigotimes_{n=1}^{P_{\rm in}}\sigma_n(\mathbf{u}_j) = \sum_m f_{m,n,j}^1(\mathbf{u}_j) \tilde\sigma_m,
\end{align}
where $\tilde\sigma_m$ are density matrices on the combined system of all input subcomponents. While linear encodings with input via re-initialization are possible, they require sophisticated mixed state engineering, as the above equation must be valid for all $j$ and any $\rho_{j-1}$.

\subsubsection{Channel mixtures}

The second example of a discrete input encoding is a mixture of fixed quantum channels applied to the quantum state at each input step, described by the transformation
\begin{align}
    \rho^{\rm in}_j = \sum_m f_{m,j}(\mathbf{u}_j)\mathcal{C}_m\left[\rho_{j-1}\right],
\end{align}
where $\mathcal{C}_m$ are the fixed quantum channels. Provided that $f_{m,j}(\mathbf{u}_j)$ are linear functions of their input parameters, then this is a linear encoding of the input. This class of input encodings was studied in Ref.~\cite{Chen:2020}, with unitary transformations for the fixed channels, and is the simplest discrete input encoding that can be made linear.

\subsubsection{Parameterized Unitaries}

As a final example, we consider a natural encoding of the input signal into the parameters of a quantum circuit, i.e.~a unitary transformation. Such an encoding was considered in Ref.~\cite{Dasgupta:2020}. In a qubit architecture, this is often done by encoding the real-valued input into the angle of a single or two-qubit gate, which results in a sinusoidal nonlinearity for the input encoding.

More generally, one can write the input encoding unitary in its diagonal basis as
\begin{align}
    \hat{U}_{\rm in}(\mathbf{u}_j) = \sum_m e^{i\varphi_m}\hat{P}_m,
\end{align}
where $\hat{P}_m$ are the projection operators for the eigenvectors of $\hat{U}$ with corresponding eigenvalues $e^{i\varphi_m} = R_{m} + iI_{m}$. The action of such a unitary is to return the state
\begin{align}
    \rho^{\rm in}_j = \sum_{m,l} e^{i(\varphi_m - \varphi_l)}\hat{P}_m\rho_{j-1}\hat{P}_l.
\end{align}
If we write each eigenvalue pair-product in terms of its real and imaginary parts as $e^{i(\varphi_m - \varphi_l)} = R_{m,l} + iI_{m,l}$, where $R_{m,l}^2 + I_{m,l}^2 = 1$, then we can achieve a linear encoding if $R_{m,l}$ and $I_{m,l}$ are linear functions of the elements of $\mathbf{u}_j$ for all $m$ and $l$.

If more than one eigenvalue, e.g.~$e^{i\varphi_n}$ and $e^{i\varphi_k}$, is a function of the input signal, then since
\begin{align}
    \nonumber&R_{n,k} + iI_{n,k} = e^{i(\varphi_n - \varphi_k)} =  e^{i\varphi_n}e^{-i\varphi_k} \\ 
    \nonumber&= (R_{n} + iI_{n})(R_{k} - iI_{k}) 
    \\ &= R_{n}R_{k} +I_nI_k + i(R_{k}I_{n} - R_{n}I_{k}), 
\end{align}
the real an imaginary components of any one eigenvalue must be nonlinear functions of the input signal parameters for their products to be linear functions. Thus, for more than one eigenvalue of the input unitary to depend on the input data we require nonlinear pre-processing. Furthermore, conversion from the physically controllable parameters in the input unitary to its eigenvalues is likely to require additional nonlinear pre-processing. 

\subsection{Continuous input encodings}

Finite dimensional systems cannot have a linear continuous input encoding as they cannot satisfy condition i). To see this, consider the adjoint master equation (in diagonal form) for the element $\hat{B}_k$ of a basis formed by Hermitian trace-zero matrices plus the identity
\begin{align}
    \partial_t \hat{B}_k = i\left[\hat{H},\hat{B}_k\right] + \sum_j \gamma_j\left( \hat{L}_j^\dagger\hat{B}_k\hat{L}_j - \frac{1}{2}\left\{\hat{L}_j^\dagger\hat{L}_j, \hat{B}_k\right\}\right) \label{eqn:adjME}.
\end{align}
For finite dimensional systems the first term on the right-hand side of Eq.~\eqref{eqn:adjME}, the communtator with the Hamiltonian, has trace zero and cannot be proportional to the identity. Similarly, the second term is also trace zero for Hermitian $\hat{B}_k$, which implies that there is no term proportional to the identity. Note that this result also follows from the fact that the adjoint master equation generates unital evolution, which implies that the identity operator is an eigen-operator of the adjoint master equation (with eigenvalue $0$).

Straightforward input encodings that encode the input signal into the parameters of a time-dependent Hamiltonian $\hat{H}(t)$ naturally implement an exponential nonlinearity on the input data \cite{Fischbacher2020}. Another noteworthy example of a nonlinear continuous input encoding for finite dimensional QRC is that of Refs.~\cite{Ghosh:2019aa,Ghosh2020_2}, which uses the cascaded interaction with an infinite dimensional ancillary system as input to the QRC. 

\section{Infinite dimensional systems}
\label{sec:infinite}

We now consider quantum systems with infinite dimensional Hilbert spaces, otherwise known as continuous variable systems. Proposals for QRC in this domain have focused on bosonic systems described by harmonic or nonlinear oscillators, and both discrete \cite{Nokkala2021} and continuous \cite{Govia2020,Kalfus2021} input schemes have been considered. For continuous variable systems, it is convenient to define the reservoir nodes as the expectation values of the basis formed by the quadrature moments
\begin{align}
    C_{nm} = \left<\hat{X}^n\hat{P}^m\right>,
\end{align}
where $\hat{X}$ and $\hat{P}$ are the position and momentum operators for a single continuous variable quantum system. The moments for $n+m = 1$ and $n+m = 2$ are the average values and covariance matrix, respectively, and form a complete description of Gaussian continuous variable systems.

\subsection{Discrete input encodings}

Just as for finite dimensional systems, infinite dimensional systems admit the linear input encodings discussed previously that either use state re-initialization into a mixed state, or a mixture of quantum channels. Beyond these, infinite dimensional systems have a straightforward linear input encoding that preserves state purity. It uses the displacement operator
\begin{align}
    \hat{D}(\beta) = \exp(\beta\hat{a}^\dagger - \beta^*\hat{a}),
\end{align}
where $\hat{a} = (\hat{X} + i\hat{P})/\sqrt{2}$ is the lowering operator and $\beta \in \mathbb{C}$, to implement the input transformation
\begin{align}
    \nonumber&\left<\hat{X}\right>_{j}  = \left<\hat{D}(\beta)\hat{X}\hat{D}^\dagger(\beta)\right>_{j-1} = \left<\hat{X}\right>_{j-1} + \sqrt{2}{\rm Re}[\beta],\\
    \nonumber&\left<\hat{P}\right>_{j} = \left<\hat{D}(\beta)\hat{P}\hat{D}^\dagger(\beta)\right>_{j-1} = \left<\hat{P}\right>_{j-1} + \sqrt{2}{\rm Im}[\beta].
\end{align}
As is clear from the above equations, linearly encoding $\mathbf{u}_j$ into the real and imaginary parts of $\beta$ results in a linear input encoding. 

Displacement operator input encoding can also be used to realize a state re-initialization input scheme that creates pure states, as proposed for Gaussian systems in Ref.~\cite{Nokkala2021}. This approach first re-initializes the target system into the vaccum state, and then applies the displacement transformation, resulting in the coherent state $\ket{\beta}$. On the other hand, a linear input encoding cannot be obtained from the squeezed state re-initialization of Ref.~\cite{Nokkala2021}, as the covariance matrix elements are different nonlinear functions of the complex squeezing parameter.

\subsection{Continuous input encodings}

Unlike the finite dimensional case, condition i) for a linear continuous input encoding \emph{can} be satisfied for infinite dimensional QRC. One way to do so, using coherent evolution, is to implement the Hamiltonian dynamics that generate a displacement transformation, as considered in Refs.~\cite{Govia2020,Kalfus2021}. As an example, for the $\hat{X}$ operator this would be
\begin{equation}
    \partial_t \hat{X} = i\left[\hat{H}(t),\hat{X}\right] = u(t)\iden,~{\rm for}~\hat{H}(t) = u(t)\hat{P},
\end{equation}
with a univariate input signal $u(t)$. Suitable choice of $u(t)$ will satisfy condition ii) of a linear continuous input encoding. It is worth noting that the short-term memory capacity benchmark of Ref.~\cite{Kalfus2021} satisfies condition ii), as it is piece-wise constant input. The sinusoidal parameter estimation tasks of Refs.~\cite{Govia2020,Kalfus2021} are linear if the signal input is encoded in amplitude, but nonlinear if it is encoded phase, which highlights how the task itself can determine linearity of the input encoding.

This linear encoding comes about due to the canonical commutation relation $\left[\hat{X},\hat{P}\right] = i$, which is only valid for an infinite dimensional system (and operators that are unitarily equivalent to $\hat{X}$ and $\hat{P}$). No other elements of the moment basis have this commutation relation, and as such linear input encoding is only possible for the quadrature operators themselves.

Expanding the discussion to consider dissipative interactions encoding the input signal, it may be possible to engineer other linear continuous input encodings for infinite dimensional systems. One possibility would be to modify the cascaded-input framework of Ref.~\cite{Ghosh:2019aa} to consider a bosonic ancilla and a bosonic reservoir. The details of such dissipative input encodings, or a proof of their nonlinearity are beyond the scope of this work.

\section{Conclusion}
\label{sec:conc}

In this manuscript, we have critically examined the input procedure in quantum reservoir computing to determine the conditions on a linear encoding of the input data. As we have shown, nonlinear input encodings are ubiquitous. Linear encodings are impossible in many proposed implementations of quantum reservoir computing, and even when possible require complex engineering to achieve. Notable exceptions are linear input encoding by a mixture of quantum channels \cite{Chen:2020}, or via a displacement transformation \cite{Govia2020,Nokkala2021,Kalfus2021}.

The relevance of this discussion stems from the fact that nonlinearity is a necessary component for nontrivial computation and task performance in reservoir computing, and thus considerable effort is spent on achieving sufficient nonlinearity in the reservoir \emph{internal} dynamics. If the input encoding is itself nonlinear, this calls into question the necessity of this effort. It may be sufficient for the internal dynamics to select the ``right'' parts of the transformed input data, as recently discussed for other frameworks of quantum machine learning \cite{Schuld20}.

For discrete input encodings, we have shown that the family of transformations that can be applied to the input data, as allowed by the laws of quantum mechanics, often result in quadratic, square-root, or exponential maps of the input data they encode. For continuous input encodings, the conditions we have derived for linear input in a quantum reservoir are more a feature of the fact that reservoir dynamics are described by an inhomogeneous differential equation than of quantum mechanics. To that end, very similar conditions would be derived for a classical reservoir. We have shown that linearity of the input parameters in this differential equation is insufficient for a linear encoding, and that the input must act like a linear ``forcing term'', which is only possible for infinite dimensional systems.

In the continuous case, since the input and internal dynamics happen simultaneously our separation of the two may seem arbitrary. However, as in the discrete case, we argue that this dichotomy is important to consider when designing reservoirs. With very limited exceptions, quantum reservoirs with continuous input impart an exponential map on the input data, which may already be sufficient for computation without complex internal dynamics.

Finally, we note that quantum reservoir computing protocols such as those in Refs.~\cite{Ghosh:2019ab,Ghosh:2020aa,Krisnanda2021} fall outside the scope of this manuscript. They have nominally constant ``input'' as their task is to prepare a resource quantum state or implement desired dynamical evolution, not to perform computation on an input of time-series data.

\acknowledgements
This material is based upon work supported by the U.S. Army Research Office under Contract No: W911NF-19-C-0092. Any opinions, findings and conclusions or recommendations expressed in this material are those of the authors and do not necessarily reflect the views of the U.S. Army Research Office.

\appendix

\section{General Description of Continuous Input QRC}
\label{app:gencon}

As described previously, the full evolution of a reservoir with continuous input is described by Eq.~\eqref{eqn:diffmapfinite}. For our purposes, it will be easier to work with the vectorized version of this equation, as this is a matrix-vector differential equation
\begin{align}
    \partial_t\superket{B(t)} = \hat{\mathcal{L}}_{A}(\mathbf{u}(t),\boldsymbol{\alpha})\superket{B(t)}, 
\end{align}
where $\superket{B(t)}$ is the vector formed by column-stacking $\hat{B}(t)$, and $\hat{\mathcal{L}}_{A}$ is the super-operator matrix representation of the linear operator $\mathcal{L}^\dagger$. Using this formalism, it is straightforward to show that the evolution for a vector of basis-operator expectation values is given by
\begin{align}
    \partial_t \mathbf{B}(t) =\hat{\mathcal{L}}_{A}(\mathbf{u}(t),\boldsymbol{\alpha})\mathbf{B}(t),
\end{align}
where $\mathbf{B}(t)$ is defined in Eq.~\eqref{eqn:expvec}, and we express $\hat{\mathcal{L}}_{A}$ in this basis as
\begin{align}
    \hat{\mathcal{L}}_{A} = \sum_{j,k} [\hat{\mathcal{L}}_{A}]_{k,j} \superket{B_j}\superbra{B_k}.
\end{align}
Now we assume condition i) for continuous input, and separate out the component of the evolution due to the input to write the evoltuion as
\begin{align}
    \partial_t \mathbf{B}(t) =\hat{\mathcal{L}}_{A}(\boldsymbol{\alpha})\mathbf{B}(t) + \mathbf{F}(\mathbf{u}(t)),
\end{align}
where $\mathbf{F}$ is the vector of functions acting on the input for each element of $\mathbf{B}(t)$.

Assuming that $\hat{\mathcal{L}}_{A}(\boldsymbol{\alpha})$ is diagonalizable, then we can make a linear transformation defined by the matrix $\hat{V}$ such that
\begin{align}
    \nonumber\partial_t \tilde{\mathbf{B}}(t) &\equiv  \partial_t \hat{V}\mathbf{B}(t) =\hat{V}\hat{\mathcal{L}}_{A}(\boldsymbol{\alpha})\hat{V}^\dagger\hat{V}\mathbf{B}(t) + \hat{V}\mathbf{F}(\mathbf{u}(t)), \\
    &= \hat{\mathcal{D}}(\boldsymbol{\alpha})\tilde{\mathbf{B}}(t) + \tilde{\mathbf{F}}(\mathbf{u}(t)).
\end{align}
Since $\hat{\mathcal{D}}$ is diagonal, each element of the above equation can be solved independently, with the general solution [analogous to Eq.~\eqref{eqn:intexplicit}]
\begin{align}
    \nonumber\left<\tilde{\hat{B}}_n(t)\right> &= \frac{1}{\epsilon_n}\left(e^{\epsilon_n(t-t_0)}\tilde{F}_n(\mathbf{u}(t_0)) - \tilde{F}_n(\mathbf{u}(t))\right) \\
    &+\frac{1}{\epsilon_n}\int_{t_0}^t ds~e^{\epsilon_n(t-s)}\frac{d}{ds}\tilde{F}_n(\mathbf{u}(t)),
\end{align}
where $\{\epsilon_n\}$ are the eigenvalues of $\hat{\mathcal{L}}_{A}(\boldsymbol{\alpha})$. From the form of the above equation it is clear that the conditions for a linear input encoding derived from Eq.~\eqref{eqn:intexplicit} hold in the general case for the elements of $\tilde{F}_n(\mathbf{u}(t))$. Importantly, because $\tilde{F}_n(\mathbf{u}(t))$ and $F_n(\mathbf{u}(t))$ are connected by a linear transformation, the linear input encoding conditions can equivalently be applied to $F_n(\mathbf{u}(t))$.

\bibliography{QRC_Input_Bib.bib}

\begin{thebibliography}{31}%
\makeatletter
\providecommand \@ifxundefined [1]{%
 \@ifx{#1\undefined}
}%
\providecommand \@ifnum [1]{%
 \ifnum #1\expandafter \@firstoftwo
 \else \expandafter \@secondoftwo
 \fi
}%
\providecommand \@ifx [1]{%
 \ifx #1\expandafter \@firstoftwo
 \else \expandafter \@secondoftwo
 \fi
}%
\providecommand \natexlab [1]{#1}%
\providecommand \enquote  [1]{``#1''}%
\providecommand \bibnamefont  [1]{#1}%
\providecommand \bibfnamefont [1]{#1}%
\providecommand \citenamefont [1]{#1}%
\providecommand \href@noop [0]{\@secondoftwo}%
\providecommand \href [0]{\begingroup \@sanitize@url \@href}%
\providecommand \@href[1]{\@@startlink{#1}\@@href}%
\providecommand \@@href[1]{\endgroup#1\@@endlink}%
\providecommand \@sanitize@url [0]{\catcode `\\12\catcode `\$12\catcode
  `\&12\catcode `\#12\catcode `\^12\catcode `\_12\catcode `\%12\relax}%
\providecommand \@@startlink[1]{}%
\providecommand \@@endlink[0]{}%
\providecommand \url  [0]{\begingroup\@sanitize@url \@url }%
\providecommand \@url [1]{\endgroup\@href {#1}{\urlprefix }}%
\providecommand \urlprefix  [0]{URL }%
\providecommand \Eprint [0]{\href }%
\providecommand \doibase [0]{https://doi.org/}%
\providecommand \selectlanguage [0]{\@gobble}%
\providecommand \bibinfo  [0]{\@secondoftwo}%
\providecommand \bibfield  [0]{\@secondoftwo}%
\providecommand \translation [1]{[#1]}%
\providecommand \BibitemOpen [0]{}%
\providecommand \bibitemStop [0]{}%
\providecommand \bibitemNoStop [0]{.\EOS\space}%
\providecommand \EOS [0]{\spacefactor3000\relax}%
\providecommand \BibitemShut  [1]{\csname bibitem#1\endcsname}%
\let\auto@bib@innerbib\@empty
\bibitem [{\citenamefont {Markovi{\'c}}\ and\ \citenamefont
  {Grollier}(2020)}]{Markovic:2020aa}%
  \BibitemOpen
  \bibfield  {author} {\bibinfo {author} {\bibfnamefont {D.}~\bibnamefont
  {Markovi{\'c}}}\ and\ \bibinfo {author} {\bibfnamefont {J.}~\bibnamefont
  {Grollier}},\ }\bibfield  {title} {\bibinfo {title} {Quantum neuromorphic
  computing},\ }\href {https://doi.org/10.1063/5.0020014} {\bibfield  {journal}
  {\bibinfo  {journal} {Applied Physics Letters}\ }\textbf {\bibinfo {volume}
  {117}},\ \bibinfo {pages} {150501} (\bibinfo {year} {2020})}\BibitemShut
  {NoStop}%
\bibitem [{\citenamefont {Mujal}\ \emph {et~al.}(2021)\citenamefont {Mujal},
  \citenamefont {Martínez-Peña}, \citenamefont {Nokkala}, \citenamefont
  {García-Beni}, \citenamefont {Giorgi}, \citenamefont {Soriano},\ and\
  \citenamefont {Zambrini}}]{Mujal21}%
  \BibitemOpen
  \bibfield  {author} {\bibinfo {author} {\bibfnamefont {P.}~\bibnamefont
  {Mujal}}, \bibinfo {author} {\bibfnamefont {R.}~\bibnamefont
  {Martínez-Peña}}, \bibinfo {author} {\bibfnamefont {J.}~\bibnamefont
  {Nokkala}}, \bibinfo {author} {\bibfnamefont {J.}~\bibnamefont
  {García-Beni}}, \bibinfo {author} {\bibfnamefont {G.~L.}\ \bibnamefont
  {Giorgi}}, \bibinfo {author} {\bibfnamefont {M.~C.}\ \bibnamefont
  {Soriano}},\ and\ \bibinfo {author} {\bibfnamefont {R.}~\bibnamefont
  {Zambrini}},\ }\href@noop {} {\bibinfo {title} {Opportunities in quantum
  reservoir computing and extreme learning machines}} (\bibinfo {year}
  {2021}),\ \Eprint {https://arxiv.org/abs/arXiv:2102.11831} {arXiv:2102.11831}
  \BibitemShut {NoStop}%
\bibitem [{\citenamefont {Maass}\ \emph {et~al.}(2002)\citenamefont {Maass},
  \citenamefont {Natschl{\"a}ger},\ and\ \citenamefont
  {Markram}}]{Maass:2002aa}%
  \BibitemOpen
  \bibfield  {author} {\bibinfo {author} {\bibfnamefont {W.}~\bibnamefont
  {Maass}}, \bibinfo {author} {\bibfnamefont {T.}~\bibnamefont
  {Natschl{\"a}ger}},\ and\ \bibinfo {author} {\bibfnamefont {H.}~\bibnamefont
  {Markram}},\ }\bibfield  {title} {\bibinfo {title} {Real-time computing
  without stable states: A new framework for neural computation based on
  perturbations},\ }\href {https://doi.org/10.1162/089976602760407955}
  {\bibfield  {journal} {\bibinfo  {journal} {Neural Computation}\ }\textbf
  {\bibinfo {volume} {14}},\ \bibinfo {pages} {2531} (\bibinfo {year}
  {2002})}\BibitemShut {NoStop}%
\bibitem [{\citenamefont {Jaeger}\ and\ \citenamefont
  {Haas}(2004)}]{Jaeger:2004aa}%
  \BibitemOpen
  \bibfield  {author} {\bibinfo {author} {\bibfnamefont {H.}~\bibnamefont
  {Jaeger}}\ and\ \bibinfo {author} {\bibfnamefont {H.}~\bibnamefont {Haas}},\
  }\bibfield  {title} {\bibinfo {title} {Harnessing nonlinearity: Predicting
  chaotic systems and saving energy in wireless communication},\ }\href
  {https://doi.org/10.1126/science.1091277} {\bibfield  {journal} {\bibinfo
  {journal} {Science}\ }\textbf {\bibinfo {volume} {304}},\ \bibinfo {pages}
  {78} (\bibinfo {year} {2004})}\BibitemShut {NoStop}%
\bibitem [{\citenamefont {Verstraeten}\ \emph {et~al.}(2007)\citenamefont
  {Verstraeten}, \citenamefont {Schrauwen}, \citenamefont {D'Haene},\ and\
  \citenamefont {Stroobandt}}]{Verstraeten:2007aa}%
  \BibitemOpen
  \bibfield  {author} {\bibinfo {author} {\bibfnamefont {D.}~\bibnamefont
  {Verstraeten}}, \bibinfo {author} {\bibfnamefont {B.}~\bibnamefont
  {Schrauwen}}, \bibinfo {author} {\bibfnamefont {M.}~\bibnamefont {D'Haene}},\
  and\ \bibinfo {author} {\bibfnamefont {D.}~\bibnamefont {Stroobandt}},\
  }\bibfield  {title} {\bibinfo {title} {An experimental unification of
  reservoir computing methods},\ }\href
  {https://doi.org/https://doi.org/10.1016/j.neunet.2007.04.003} {\bibfield
  {journal} {\bibinfo  {journal} {Neural Networks}\ }\textbf {\bibinfo {volume}
  {20}},\ \bibinfo {pages} {391 } (\bibinfo {year} {2007})}\BibitemShut
  {NoStop}%
\bibitem [{\citenamefont {Tanaka}\ \emph {et~al.}(2019)\citenamefont {Tanaka},
  \citenamefont {Yamane}, \citenamefont {H{\'e}roux}, \citenamefont {Nakane},
  \citenamefont {Kanazawa}, \citenamefont {Takeda}, \citenamefont {Numata},
  \citenamefont {Nakano},\ and\ \citenamefont {Hirose}}]{Tanaka:2019aa}%
  \BibitemOpen
  \bibfield  {author} {\bibinfo {author} {\bibfnamefont {G.}~\bibnamefont
  {Tanaka}}, \bibinfo {author} {\bibfnamefont {T.}~\bibnamefont {Yamane}},
  \bibinfo {author} {\bibfnamefont {J.~B.}\ \bibnamefont {H{\'e}roux}},
  \bibinfo {author} {\bibfnamefont {R.}~\bibnamefont {Nakane}}, \bibinfo
  {author} {\bibfnamefont {N.}~\bibnamefont {Kanazawa}}, \bibinfo {author}
  {\bibfnamefont {S.}~\bibnamefont {Takeda}}, \bibinfo {author} {\bibfnamefont
  {H.}~\bibnamefont {Numata}}, \bibinfo {author} {\bibfnamefont
  {D.}~\bibnamefont {Nakano}},\ and\ \bibinfo {author} {\bibfnamefont
  {A.}~\bibnamefont {Hirose}},\ }\bibfield  {title} {\bibinfo {title} {Recent
  advances in physical reservoir computing: A review},\ }\href
  {https://doi.org/https://doi.org/10.1016/j.neunet.2019.03.005} {\bibfield
  {journal} {\bibinfo  {journal} {Neural Networks}\ }\textbf {\bibinfo {volume}
  {115}},\ \bibinfo {pages} {100 } (\bibinfo {year} {2019})}\BibitemShut
  {NoStop}%
\bibitem [{\citenamefont {Abreu~Araujo}\ \emph {et~al.}(2020)\citenamefont
  {Abreu~Araujo}, \citenamefont {Riou}, \citenamefont {Torrejon}, \citenamefont
  {Tsunegi}, \citenamefont {Querlioz}, \citenamefont {Yakushiji}, \citenamefont
  {Fukushima}, \citenamefont {Kubota}, \citenamefont {Yuasa}, \citenamefont
  {Stiles},\ and\ \citenamefont {Grollier}}]{Abreu-Araujo:2020wy}%
  \BibitemOpen
  \bibfield  {author} {\bibinfo {author} {\bibfnamefont {F.}~\bibnamefont
  {Abreu~Araujo}}, \bibinfo {author} {\bibfnamefont {M.}~\bibnamefont {Riou}},
  \bibinfo {author} {\bibfnamefont {J.}~\bibnamefont {Torrejon}}, \bibinfo
  {author} {\bibfnamefont {S.}~\bibnamefont {Tsunegi}}, \bibinfo {author}
  {\bibfnamefont {D.}~\bibnamefont {Querlioz}}, \bibinfo {author}
  {\bibfnamefont {K.}~\bibnamefont {Yakushiji}}, \bibinfo {author}
  {\bibfnamefont {A.}~\bibnamefont {Fukushima}}, \bibinfo {author}
  {\bibfnamefont {H.}~\bibnamefont {Kubota}}, \bibinfo {author} {\bibfnamefont
  {S.}~\bibnamefont {Yuasa}}, \bibinfo {author} {\bibfnamefont {M.~D.}\
  \bibnamefont {Stiles}},\ and\ \bibinfo {author} {\bibfnamefont
  {J.}~\bibnamefont {Grollier}},\ }\bibfield  {title} {\bibinfo {title} {Role
  of non-linear data processing on speech recognition task in the framework of
  reservoir computing},\ }\href {https://doi.org/10.1038/s41598-019-56991-x}
  {\bibfield  {journal} {\bibinfo  {journal} {Scientific Reports}\ }\textbf
  {\bibinfo {volume} {10}},\ \bibinfo {pages} {328} (\bibinfo {year}
  {2020})}\BibitemShut {NoStop}%
\bibitem [{\citenamefont {Chen}\ and\ \citenamefont
  {Nurdin}(2019)}]{Chen:2019aa}%
  \BibitemOpen
  \bibfield  {author} {\bibinfo {author} {\bibfnamefont {J.}~\bibnamefont
  {Chen}}\ and\ \bibinfo {author} {\bibfnamefont {H.~I.}\ \bibnamefont
  {Nurdin}},\ }\bibfield  {title} {\bibinfo {title} {Learning nonlinear
  input--output maps with dissipative quantum systems},\ }\href
  {https://doi.org/10.1007/s11128-019-2311-9} {\bibfield  {journal} {\bibinfo
  {journal} {Quantum Information Processing}\ }\textbf {\bibinfo {volume}
  {18}},\ \bibinfo {pages} {198} (\bibinfo {year} {2019})}\BibitemShut
  {NoStop}%
\bibitem [{\citenamefont {Magnus}(1954)}]{Magnus1954}%
  \BibitemOpen
  \bibfield  {author} {\bibinfo {author} {\bibfnamefont {W.}~\bibnamefont
  {Magnus}},\ }\bibfield  {title} {\bibinfo {title} {{On the exponential
  solution of differential equations for a linear operator}},\ }\href
  {https://doi.org/10.1002/cpa.3160070404} {\bibfield  {journal} {\bibinfo
  {journal} {{Comm. Pure Appl. Math.}}\ }\textbf {\bibinfo {volume} {7}},\
  \bibinfo {pages} {649} (\bibinfo {year} {1954})}\BibitemShut {NoStop}%
\bibitem [{\citenamefont {Fujii}\ and\ \citenamefont
  {Nakajima}(2017)}]{Fujii:2017aa}%
  \BibitemOpen
  \bibfield  {author} {\bibinfo {author} {\bibfnamefont {K.}~\bibnamefont
  {Fujii}}\ and\ \bibinfo {author} {\bibfnamefont {K.}~\bibnamefont
  {Nakajima}},\ }\bibfield  {title} {\bibinfo {title} {{Harnessing
  Disordered-Ensemble Quantum Dynamics for Machine Learning}},\ }\href
  {https://doi.org/10.1103/PhysRevApplied.8.024030} {\bibfield  {journal}
  {\bibinfo  {journal} {Phys. Rev. Applied}\ }\textbf {\bibinfo {volume} {8}},\
  \bibinfo {pages} {24030} (\bibinfo {year} {2017})}\BibitemShut {NoStop}%
\bibitem [{\citenamefont {Nakajima}\ \emph {et~al.}(2019)\citenamefont
  {Nakajima}, \citenamefont {Fujii}, \citenamefont {Negoro}, \citenamefont
  {Mitarai},\ and\ \citenamefont {Kitagawa}}]{Nakajima:2019aa}%
  \BibitemOpen
  \bibfield  {author} {\bibinfo {author} {\bibfnamefont {K.}~\bibnamefont
  {Nakajima}}, \bibinfo {author} {\bibfnamefont {K.}~\bibnamefont {Fujii}},
  \bibinfo {author} {\bibfnamefont {M.}~\bibnamefont {Negoro}}, \bibinfo
  {author} {\bibfnamefont {K.}~\bibnamefont {Mitarai}},\ and\ \bibinfo {author}
  {\bibfnamefont {M.}~\bibnamefont {Kitagawa}},\ }\bibfield  {title} {\bibinfo
  {title} {{Boosting Computational Power through Spatial Multiplexing in
  Quantum Reservoir Computing}},\ }\href
  {https://doi.org/10.1103/PhysRevApplied.11.034021} {\bibfield  {journal}
  {\bibinfo  {journal} {Phys. Rev. Applied}\ }\textbf {\bibinfo {volume}
  {11}},\ \bibinfo {pages} {34021} (\bibinfo {year} {2019})}\BibitemShut
  {NoStop}%
\bibitem [{\citenamefont {Negoro}\ \emph {et~al.}(2018)\citenamefont {Negoro},
  \citenamefont {Mitarai}, \citenamefont {Fujii}, \citenamefont {Nakajima},\
  and\ \citenamefont {Kitagawa}}]{Negoro:2018aa}%
  \BibitemOpen
  \bibfield  {author} {\bibinfo {author} {\bibfnamefont {M.}~\bibnamefont
  {Negoro}}, \bibinfo {author} {\bibfnamefont {K.}~\bibnamefont {Mitarai}},
  \bibinfo {author} {\bibfnamefont {K.}~\bibnamefont {Fujii}}, \bibinfo
  {author} {\bibfnamefont {K.}~\bibnamefont {Nakajima}},\ and\ \bibinfo
  {author} {\bibfnamefont {M.}~\bibnamefont {Kitagawa}},\ }\href@noop {}
  {\bibinfo {title} {Machine learning with controllable quantum dynamics of a
  nuclear spin ensemble in a solid}} (\bibinfo {year} {2018}),\ \Eprint
  {https://arxiv.org/abs/arXiv:1806.10910} {arXiv:1806.10910} \BibitemShut
  {NoStop}%
\bibitem [{\citenamefont {Kutvonen}\ \emph {et~al.}(2020)\citenamefont
  {Kutvonen}, \citenamefont {Fujii},\ and\ \citenamefont
  {Sagawa}}]{Kutvonen:2020aa}%
  \BibitemOpen
  \bibfield  {author} {\bibinfo {author} {\bibfnamefont {A.}~\bibnamefont
  {Kutvonen}}, \bibinfo {author} {\bibfnamefont {K.}~\bibnamefont {Fujii}},\
  and\ \bibinfo {author} {\bibfnamefont {T.}~\bibnamefont {Sagawa}},\
  }\bibfield  {title} {\bibinfo {title} {Optimizing a quantum reservoir
  computer for time series prediction},\ }\href
  {https://doi.org/10.1038/s41598-020-71673-9} {\bibfield  {journal} {\bibinfo
  {journal} {Scientific Reports}\ }\textbf {\bibinfo {volume} {10}},\ \bibinfo
  {pages} {14687} (\bibinfo {year} {2020})}\BibitemShut {NoStop}%
\bibitem [{\citenamefont {Mart{\'\i}nez-Pe{\~n}a}\ \emph
  {et~al.}(2020)\citenamefont {Mart{\'\i}nez-Pe{\~n}a}, \citenamefont
  {Nokkala}, \citenamefont {Giorgi}, \citenamefont {Zambrini},\ and\
  \citenamefont {Soriano}}]{Martinez-Pena:2020ts}%
  \BibitemOpen
  \bibfield  {author} {\bibinfo {author} {\bibfnamefont {R.}~\bibnamefont
  {Mart{\'\i}nez-Pe{\~n}a}}, \bibinfo {author} {\bibfnamefont {J.}~\bibnamefont
  {Nokkala}}, \bibinfo {author} {\bibfnamefont {G.~L.}\ \bibnamefont {Giorgi}},
  \bibinfo {author} {\bibfnamefont {R.}~\bibnamefont {Zambrini}},\ and\
  \bibinfo {author} {\bibfnamefont {M.~C.}\ \bibnamefont {Soriano}},\
  }\bibfield  {title} {\bibinfo {title} {Information processing capacity of
  spin-based quantum reservoir computing systems},\ }\bibfield  {journal}
  {\bibinfo  {journal} {Cognitive Computation}\ }\href
  {https://doi.org/10.1007/s12559-020-09772-y} {10.1007/s12559-020-09772-y}
  (\bibinfo {year} {2020})\BibitemShut {NoStop}%
\bibitem [{\citenamefont {Fujii}\ and\ \citenamefont
  {Nakajima}(2020)}]{Fujii2020}%
  \BibitemOpen
  \bibfield  {author} {\bibinfo {author} {\bibfnamefont {K.}~\bibnamefont
  {Fujii}}\ and\ \bibinfo {author} {\bibfnamefont {K.}~\bibnamefont
  {Nakajima}},\ }\href@noop {} {\bibinfo {title} {Quantum reservoir computing:
  a reservoir approach toward quantum machine learning on near-term quantum
  devices}} (\bibinfo {year} {2020}),\ \Eprint
  {https://arxiv.org/abs/arXiv:2011.04890} {arXiv:2011.04890} \BibitemShut
  {NoStop}%
\bibitem [{\citenamefont {Martínez-Peña}\ \emph {et~al.}(2021)\citenamefont
  {Martínez-Peña}, \citenamefont {Giorgi}, \citenamefont {Nokkala},
  \citenamefont {Soriano},\ and\ \citenamefont {Zambrini}}]{Pena2021}%
  \BibitemOpen
  \bibfield  {author} {\bibinfo {author} {\bibfnamefont {R.}~\bibnamefont
  {Martínez-Peña}}, \bibinfo {author} {\bibfnamefont {G.~L.}\ \bibnamefont
  {Giorgi}}, \bibinfo {author} {\bibfnamefont {J.}~\bibnamefont {Nokkala}},
  \bibinfo {author} {\bibfnamefont {M.~C.}\ \bibnamefont {Soriano}},\ and\
  \bibinfo {author} {\bibfnamefont {R.}~\bibnamefont {Zambrini}},\ }\href@noop
  {} {\bibinfo {title} {Dynamical phase transitions in quantum reservoir
  computing}} (\bibinfo {year} {2021}),\ \Eprint
  {https://arxiv.org/abs/arXiv:2103.05348} {arXiv:2103.05348} \BibitemShut
  {NoStop}%
\bibitem [{\citenamefont {Xia}\ \emph {et~al.}(2021)\citenamefont {Xia},
  \citenamefont {Zou}, \citenamefont {Qiu},\ and\ \citenamefont
  {Li}}]{Xia2021}%
  \BibitemOpen
  \bibfield  {author} {\bibinfo {author} {\bibfnamefont {W.}~\bibnamefont
  {Xia}}, \bibinfo {author} {\bibfnamefont {J.}~\bibnamefont {Zou}}, \bibinfo
  {author} {\bibfnamefont {X.}~\bibnamefont {Qiu}},\ and\ \bibinfo {author}
  {\bibfnamefont {X.}~\bibnamefont {Li}},\ }\href@noop {} {\bibinfo {title}
  {The reservoir learning power across quantum many-boby localization
  transition}} (\bibinfo {year} {2021}),\ \Eprint
  {https://arxiv.org/abs/arXiv:2104.02727} {arXiv:2104.02727} \BibitemShut
  {NoStop}%
\bibitem [{\citenamefont {Tran}\ and\ \citenamefont
  {Nakajima}(2021)}]{Tran2021}%
  \BibitemOpen
  \bibfield  {author} {\bibinfo {author} {\bibfnamefont {Q.~H.}\ \bibnamefont
  {Tran}}\ and\ \bibinfo {author} {\bibfnamefont {K.}~\bibnamefont
  {Nakajima}},\ }\href@noop {} {\bibinfo {title} {Learning temporal quantum
  tomography}} (\bibinfo {year} {2021}),\ \Eprint
  {https://arxiv.org/abs/arXiv:2103.13973} {arXiv:2103.13973} \BibitemShut
  {NoStop}%
\bibitem [{\citenamefont {Tran}\ and\ \citenamefont
  {Nakajima}(2020)}]{Tran2020}%
  \BibitemOpen
  \bibfield  {author} {\bibinfo {author} {\bibfnamefont {Q.~H.}\ \bibnamefont
  {Tran}}\ and\ \bibinfo {author} {\bibfnamefont {K.}~\bibnamefont
  {Nakajima}},\ }\href@noop {} {\bibinfo {title} {Higher-order quantum
  reservoir computing}} (\bibinfo {year} {2020}),\ \Eprint
  {https://arxiv.org/abs/arXiv:2006.08999} {arXiv:2006.08999} \BibitemShut
  {NoStop}%
\bibitem [{\citenamefont {Chen}\ \emph {et~al.}(2020)\citenamefont {Chen},
  \citenamefont {Nurdin},\ and\ \citenamefont {Yamamoto}}]{Chen:2020}%
  \BibitemOpen
  \bibfield  {author} {\bibinfo {author} {\bibfnamefont {J.}~\bibnamefont
  {Chen}}, \bibinfo {author} {\bibfnamefont {H.~I.}\ \bibnamefont {Nurdin}},\
  and\ \bibinfo {author} {\bibfnamefont {N.}~\bibnamefont {Yamamoto}},\
  }\bibfield  {title} {\bibinfo {title} {Temporal information processing on
  noisy quantum computers},\ }\href
  {https://doi.org/10.1103/PhysRevApplied.14.024065} {\bibfield  {journal}
  {\bibinfo  {journal} {Phys. Rev. Applied}\ }\textbf {\bibinfo {volume}
  {14}},\ \bibinfo {pages} {024065} (\bibinfo {year} {2020})}\BibitemShut
  {NoStop}%
\bibitem [{\citenamefont {Dasgupta}\ \emph {et~al.}(2020)\citenamefont
  {Dasgupta}, \citenamefont {Hamilton}, \citenamefont {Lougovski},\ and\
  \citenamefont {Banerjee}}]{Dasgupta:2020}%
  \BibitemOpen
  \bibfield  {author} {\bibinfo {author} {\bibfnamefont {S.}~\bibnamefont
  {Dasgupta}}, \bibinfo {author} {\bibfnamefont {K.~E.}\ \bibnamefont
  {Hamilton}}, \bibinfo {author} {\bibfnamefont {P.}~\bibnamefont
  {Lougovski}},\ and\ \bibinfo {author} {\bibfnamefont {A.}~\bibnamefont
  {Banerjee}},\ }\href@noop {} {\bibinfo {title} {Designing a nisq reservoir
  with maximal memory capacity for volatility forecasting}} (\bibinfo {year}
  {2020}),\ \Eprint {https://arxiv.org/abs/arXiv:2004.08240} {arXiv:2004.08240}
  \BibitemShut {NoStop}%
\bibitem [{\citenamefont {Fischbacher}\ \emph {et~al.}(2020)\citenamefont
  {Fischbacher}, \citenamefont {Comsa}, \citenamefont {Potempa}, \citenamefont
  {Firsching}, \citenamefont {Versari},\ and\ \citenamefont
  {Alakuijala}}]{Fischbacher2020}%
  \BibitemOpen
  \bibfield  {author} {\bibinfo {author} {\bibfnamefont {T.}~\bibnamefont
  {Fischbacher}}, \bibinfo {author} {\bibfnamefont {I.~M.}\ \bibnamefont
  {Comsa}}, \bibinfo {author} {\bibfnamefont {K.}~\bibnamefont {Potempa}},
  \bibinfo {author} {\bibfnamefont {M.}~\bibnamefont {Firsching}}, \bibinfo
  {author} {\bibfnamefont {L.}~\bibnamefont {Versari}},\ and\ \bibinfo {author}
  {\bibfnamefont {J.}~\bibnamefont {Alakuijala}},\ }\href@noop {} {\bibinfo
  {title} {Intelligent matrix exponentiation}} (\bibinfo {year} {2020}),\
  \Eprint {https://arxiv.org/abs/arXiv:2008.03936} {arXiv:2008.03936}
  \BibitemShut {NoStop}%
\bibitem [{\citenamefont {Ghosh}\ \emph
  {et~al.}(2019{\natexlab{a}})\citenamefont {Ghosh}, \citenamefont {Opala},
  \citenamefont {Matuszewski}, \citenamefont {Paterek},\ and\ \citenamefont
  {Liew}}]{Ghosh:2019aa}%
  \BibitemOpen
  \bibfield  {author} {\bibinfo {author} {\bibfnamefont {S.}~\bibnamefont
  {Ghosh}}, \bibinfo {author} {\bibfnamefont {A.}~\bibnamefont {Opala}},
  \bibinfo {author} {\bibfnamefont {M.}~\bibnamefont {Matuszewski}}, \bibinfo
  {author} {\bibfnamefont {T.}~\bibnamefont {Paterek}},\ and\ \bibinfo {author}
  {\bibfnamefont {T.~C.~H.}\ \bibnamefont {Liew}},\ }\bibfield  {title}
  {\bibinfo {title} {{Quantum reservoir processing}},\ }\href
  {https://doi.org/10.1038/s41534-019-0149-8} {\bibfield  {journal} {\bibinfo
  {journal} {npj Quantum Information}\ }\textbf {\bibinfo {volume} {5}}
  (\bibinfo {year} {2019}{\natexlab{a}})}\BibitemShut {NoStop}%
\bibitem [{\citenamefont {Ghosh}\ \emph
  {et~al.}(2020{\natexlab{a}})\citenamefont {Ghosh}, \citenamefont {Opala},
  \citenamefont {Matuszewski}, \citenamefont {Paterek},\ and\ \citenamefont
  {Liew}}]{Ghosh2020_2}%
  \BibitemOpen
  \bibfield  {author} {\bibinfo {author} {\bibfnamefont {S.}~\bibnamefont
  {Ghosh}}, \bibinfo {author} {\bibfnamefont {A.}~\bibnamefont {Opala}},
  \bibinfo {author} {\bibfnamefont {M.}~\bibnamefont {Matuszewski}}, \bibinfo
  {author} {\bibfnamefont {T.}~\bibnamefont {Paterek}},\ and\ \bibinfo {author}
  {\bibfnamefont {T.~C.~H.}\ \bibnamefont {Liew}},\ }\bibfield  {title}
  {\bibinfo {title} {Reconstructing quantum states with quantum reservoir
  networks},\ }\href {https://doi.org/10.1109/TNNLS.2020.3009716} {\bibfield
  {journal} {\bibinfo  {journal} {IEEE Trans. Neural Netw. Learn. Syst.}\ ,\
  \bibinfo {pages} {1}} (\bibinfo {year} {2020}{\natexlab{a}})}\BibitemShut
  {NoStop}%
\bibitem [{\citenamefont {Nokkala}\ \emph {et~al.}(2021)\citenamefont
  {Nokkala}, \citenamefont {Mart{\'\i}nez-Pe{\~n}a}, \citenamefont {Giorgi},
  \citenamefont {Parigi}, \citenamefont {Soriano},\ and\ \citenamefont
  {Zambrini}}]{Nokkala2021}%
  \BibitemOpen
  \bibfield  {author} {\bibinfo {author} {\bibfnamefont {J.}~\bibnamefont
  {Nokkala}}, \bibinfo {author} {\bibfnamefont {R.}~\bibnamefont
  {Mart{\'\i}nez-Pe{\~n}a}}, \bibinfo {author} {\bibfnamefont {G.~L.}\
  \bibnamefont {Giorgi}}, \bibinfo {author} {\bibfnamefont {V.}~\bibnamefont
  {Parigi}}, \bibinfo {author} {\bibfnamefont {M.~C.}\ \bibnamefont
  {Soriano}},\ and\ \bibinfo {author} {\bibfnamefont {R.}~\bibnamefont
  {Zambrini}},\ }\bibfield  {title} {\bibinfo {title} {Gaussian states of
  continuous-variable quantum systems provide universal and versatile reservoir
  computing},\ }\href {https://doi.org/10.1038/s42005-021-00556-w} {\bibfield
  {journal} {\bibinfo  {journal} {Commun Phys}\ }\textbf {\bibinfo {volume}
  {4}},\ \bibinfo {pages} {53} (\bibinfo {year} {2021})}\BibitemShut {NoStop}%
\bibitem [{\citenamefont {Govia}\ \emph {et~al.}(2021)\citenamefont {Govia},
  \citenamefont {Ribeill}, \citenamefont {Rowlands}, \citenamefont {Krovi},\
  and\ \citenamefont {Ohki}}]{Govia2020}%
  \BibitemOpen
  \bibfield  {author} {\bibinfo {author} {\bibfnamefont {L.~C.~G.}\
  \bibnamefont {Govia}}, \bibinfo {author} {\bibfnamefont {G.~J.}\ \bibnamefont
  {Ribeill}}, \bibinfo {author} {\bibfnamefont {G.~E.}\ \bibnamefont
  {Rowlands}}, \bibinfo {author} {\bibfnamefont {H.~K.}\ \bibnamefont
  {Krovi}},\ and\ \bibinfo {author} {\bibfnamefont {T.~A.}\ \bibnamefont
  {Ohki}},\ }\bibfield  {title} {\bibinfo {title} {Quantum reservoir computing
  with a single nonlinear oscillator},\ }\href
  {https://doi.org/10.1103/PhysRevResearch.3.013077} {\bibfield  {journal}
  {\bibinfo  {journal} {Phys. Rev. Research}\ }\textbf {\bibinfo {volume}
  {3}},\ \bibinfo {pages} {013077} (\bibinfo {year} {2021})}\BibitemShut
  {NoStop}%
\bibitem [{\citenamefont {Kalfus}\ \emph {et~al.}(2021)\citenamefont {Kalfus},
  \citenamefont {Ribeill}, \citenamefont {Rowlands}, \citenamefont {Krovi},
  \citenamefont {Ohki},\ and\ \citenamefont {Govia}}]{Kalfus2021}%
  \BibitemOpen
  \bibfield  {author} {\bibinfo {author} {\bibfnamefont {W.~D.}\ \bibnamefont
  {Kalfus}}, \bibinfo {author} {\bibfnamefont {G.~J.}\ \bibnamefont {Ribeill}},
  \bibinfo {author} {\bibfnamefont {G.~E.}\ \bibnamefont {Rowlands}}, \bibinfo
  {author} {\bibfnamefont {H.~K.}\ \bibnamefont {Krovi}}, \bibinfo {author}
  {\bibfnamefont {T.~A.}\ \bibnamefont {Ohki}},\ and\ \bibinfo {author}
  {\bibfnamefont {L.~C.~G.}\ \bibnamefont {Govia}},\ }\href@noop {} {\bibinfo
  {title} {Neuromorphic computing with a single qudit}} (\bibinfo {year}
  {2021}),\ \Eprint {https://arxiv.org/abs/arXiv:2101.11729} {arXiv:2101.11729}
  \BibitemShut {NoStop}%
\bibitem [{\citenamefont {Schuld}\ \emph {et~al.}(2021)\citenamefont {Schuld},
  \citenamefont {Sweke},\ and\ \citenamefont {Meyer}}]{Schuld20}%
  \BibitemOpen
  \bibfield  {author} {\bibinfo {author} {\bibfnamefont {M.}~\bibnamefont
  {Schuld}}, \bibinfo {author} {\bibfnamefont {R.}~\bibnamefont {Sweke}},\ and\
  \bibinfo {author} {\bibfnamefont {J.~J.}\ \bibnamefont {Meyer}},\ }\bibfield
  {title} {\bibinfo {title} {Effect of data encoding on the expressive power of
  variational quantum-machine-learning models},\ }\href
  {https://doi.org/10.1103/PhysRevA.103.032430} {\bibfield  {journal} {\bibinfo
   {journal} {Phys. Rev. A}\ }\textbf {\bibinfo {volume} {103}},\ \bibinfo
  {pages} {032430} (\bibinfo {year} {2021})}\BibitemShut {NoStop}%
\bibitem [{\citenamefont {Ghosh}\ \emph
  {et~al.}(2019{\natexlab{b}})\citenamefont {Ghosh}, \citenamefont {Paterek},\
  and\ \citenamefont {Liew}}]{Ghosh:2019ab}%
  \BibitemOpen
  \bibfield  {author} {\bibinfo {author} {\bibfnamefont {S.}~\bibnamefont
  {Ghosh}}, \bibinfo {author} {\bibfnamefont {T.}~\bibnamefont {Paterek}},\
  and\ \bibinfo {author} {\bibfnamefont {T.~C.~H.}\ \bibnamefont {Liew}},\
  }\bibfield  {title} {\bibinfo {title} {Quantum neuromorphic platform for
  quantum state preparation},\ }\href
  {https://doi.org/10.1103/PhysRevLett.123.260404} {\bibfield  {journal}
  {\bibinfo  {journal} {Phys. Rev. Lett.}\ }\textbf {\bibinfo {volume} {123}},\
  \bibinfo {pages} {260404} (\bibinfo {year} {2019}{\natexlab{b}})}\BibitemShut
  {NoStop}%
\bibitem [{\citenamefont {Ghosh}\ \emph
  {et~al.}(2020{\natexlab{b}})\citenamefont {Ghosh}, \citenamefont {Krisnanda},
  \citenamefont {Paterek},\ and\ \citenamefont {Liew}}]{Ghosh:2020aa}%
  \BibitemOpen
  \bibfield  {author} {\bibinfo {author} {\bibfnamefont {S.}~\bibnamefont
  {Ghosh}}, \bibinfo {author} {\bibfnamefont {T.}~\bibnamefont {Krisnanda}},
  \bibinfo {author} {\bibfnamefont {T.}~\bibnamefont {Paterek}},\ and\ \bibinfo
  {author} {\bibfnamefont {T.~C.~H.}\ \bibnamefont {Liew}},\ }\href@noop {}
  {\bibinfo {title} {Universal quantum reservoir computing}} (\bibinfo {year}
  {2020}{\natexlab{b}}),\ \Eprint {https://arxiv.org/abs/arXiv:2003.09569}
  {arXiv:2003.09569} \BibitemShut {NoStop}%
\bibitem [{\citenamefont {Krisnanda}\ \emph {et~al.}(2021)\citenamefont
  {Krisnanda}, \citenamefont {Ghosh}, \citenamefont {Paterek},\ and\
  \citenamefont {Liew}}]{Krisnanda2021}%
  \BibitemOpen
  \bibfield  {author} {\bibinfo {author} {\bibfnamefont {T.}~\bibnamefont
  {Krisnanda}}, \bibinfo {author} {\bibfnamefont {S.}~\bibnamefont {Ghosh}},
  \bibinfo {author} {\bibfnamefont {T.}~\bibnamefont {Paterek}},\ and\ \bibinfo
  {author} {\bibfnamefont {T.~C.}\ \bibnamefont {Liew}},\ }\bibfield  {title}
  {\bibinfo {title} {Creating and concentrating quantum resource states in
  noisy environments using a quantum neural network},\ }\href
  {https://doi.org/https://doi.org/10.1016/j.neunet.2021.01.003} {\bibfield
  {journal} {\bibinfo  {journal} {Neural Networks}\ }\textbf {\bibinfo {volume}
  {136}},\ \bibinfo {pages} {141} (\bibinfo {year} {2021})}\BibitemShut
  {NoStop}%
\end{thebibliography}%
\end{document}